\documentclass[]{iopart}

\usepackage{amsfonts}
\usepackage{amssymb}
\usepackage{graphicx}
\usepackage{moreverb}
\usepackage{url}
\usepackage{epsf}
\usepackage{color}
\usepackage{multirow}
\usepackage{subfigure}

\makeatletter
\let\gplgaddtomacro\g@addto@macro
\makeatother
\newcommand{\SZ}{$\mathrm{s}\zeta$}
\newcommand{\DZPmv}{$^{(P)}\mathrm{d}\zeta+\mathrm{p}$}
\newcommand{\DZP}{$\mathrm{d}\zeta+\mathrm{p}$}
\newcommand{\TZPmv}{$^{(P)}\mathrm{t}\zeta+\mathrm{p}$}
\newcommand{\TZP}{$\mathrm{t}\zeta+\mathrm{p}$}
\newcommand{\TZDPH}{$\mathrm{t}\zeta+\mathrm{d_{(H)}p}$}
\newcommand{\TZDP}{$\mathrm{t}\zeta+\mathrm{dp}$}
\newcommand{\TZDPP}{$\mathrm{t}\zeta+\mathrm{dp}+\mathrm{p}'$}
\newcommand{\QZDPs}{$^{(\mathrm{s})}\mathrm{q}\zeta+\mathrm{dp}$}
\newcommand{\QZDP}{$\mathrm{q}\zeta+\mathrm{dp}$}
\newcommand{\QZDPP}{$\mathrm{q}\zeta+\mathrm{dp}+\mathrm{p}'$}
\newcommand{\QZDPPDO}{$\mathrm{q}\zeta+\mathrm{dp}+\mathrm{p}'+\mathrm{s_{(O)}d}$}
\newcommand{\QZTPHP}{$\mathrm{q}\zeta+\mathrm{t_{(H)}p}+\mathrm{p}'$}
\newcommand{\QZTPHDPH}{$\mathrm{q}\zeta+\mathrm{t_{(H)}p}+\mathrm{d_{(H)}p}'$}
\newcommand{\QZTPDPH}{$\mathrm{q}\zeta+\mathrm{tp}+\mathrm{d_{(H)}p}'$}
\newcommand{\QZTPDP}{$\mathrm{q}\zeta+\mathrm{tp}+\mathrm{dp}'$}
\newcommand{\QZTPDPP}{$\mathrm{q}\zeta+\mathrm{tp}+\mathrm{dp}'+\mathrm{p}''$}

\begin{document}

\title{Optimal finite-range atomic basis sets for liquid water and ice}
\author{Fabiano~Corsetti$^1$, M-V~Fern\'{a}ndez-Serra$^2$, Jos\'{e}~M~Soler$^3$ and Emilio~Artacho$^{1,4,5,6}$}
\address{$^1$ CIC nanoGUNE, 20018 Donostia-San Sebasti\'{a}n, Spain}
\address{$^2$ Dept. of Physics and Astronomy, Stony Brook University, Stony Brook, New York 11794-3800, USA}
\address{$^3$ Dept. de F\'{i}sica de la Materia Condensada, Universidad Aut\'{o}noma de Madrid, 28049 Madrid, Spain}
\address{$^4$ Theory of Condensed Matter, Cavendish Laboratory, University of Cambridge, Cambridge CB3 0HE, United Kingdom}
\address{$^5$ Basque Foundation for Science Ikerbasque, 48011 Bilbao, Spain}
\address{$^6$ Donostia International Physics Center, 20018 Donostia-San Sebasti\'{a}n, Spain}
\eads{f.corsetti@nanogune.eu}

\begin{abstract}
Finite-range numerical atomic orbitals are the basis functions of choice for several first principles methods, due to their flexibility and scalability. Generating and testing such basis sets, however, remains a significant challenge for the end user. We discuss these issues and present a new scheme for generating improved polarization orbitals of finite range. We then develop a series of high-accuracy basis sets for the water molecule, and report on their performance in describing the monomer and dimer, two phases of ice, and liquid water at ambient and high density. The tests are performed by comparison with plane-wave calculations, and show the atomic orbital basis sets to exhibit an excellent level of transferability and consistency. The highest-order bases (quadruple-$\zeta$) are shown to give accuracies comparable to a plane-wave kinetic energy cutoff of at least $\sim$1000~eV for quantities such as energy differences and ionic forces, as well as achieving significantly greater accuracies for total energies and absolute pressures.
\end{abstract}

\maketitle

\section{Introduction}

Many large-scale density-functional theory (DFT) methods make use of localized basis functions, since locality in some form is an essential prerequisite for developing a method that scales linearly with system size~\cite{Bowler2012}. In particular, the use of strictly localized orbitals~\cite{Sankey1989} results in the Hamiltonian being formally sparse, without needing to impose a cutoff tolerance on the matrix elements.

Numerical atomic orbitals (NAOs) of finite range~\cite{Sankey1989} have been found to be particularly well-suited for DFT calculations, due to the fact that they are very flexible, and, therefore, only a small number of them is usually needed to obtain accurate results~\cite{Kenny2000,Junquera2001,Anglada2002,Ozaki2003,Ozaki2004,Garcia-Gil2009}. The transferability of the NAO basis between different systems has also been found to be quite reasonable~\cite{Kenny2000,Junquera2001}.

The most important drawback of using NAOs is the lack of a systematic way of improving the basis to arbitrary precision, such as what can easily be achieved in plane-wave (PW) methods~\cite{Payne1992} by increasing the kinetic energy cutoff, or in real-space-grid methods~\cite{Chelikowsky1994,Briggs1995} by decreasing the grid spacing. This is an important shortcoming of the method, since it means that it is not possible to assess the accuracy of the calculation with respect to the basis set with any degree of certainty. In contrast, it is standard practice in studies using PW methods to perform preliminary convergence tests for the kinetic energy cutoff on representative samples of the system in question, in order to tune the cutoff to be used in the production calculations to a given degree of accuracy.

In this paper, we aim to demonstrate best practice in the generation of transferable NAO basis sets of increasing accuracy for ordered and disordered bulk material, and to show how it is possible to determine with confidence the level of accuracy of the basis by careful comparison with PW calculations.

As a test case, we focus on water, in the liquid and two solid phases. Water is a material of fundamental importance in many different areas of science, and is currently of great interest to practitioners of {\em ab initio} techniques, since the subtle interplay between different interaction mechanisms that gives rise to its unusual properties (e.g., its notoriously complex phase diagram~\cite{Chen2011}, and the large number of anomalies in the liquid phase~\cite{Robinson1996,Errington2001}) is still not well understood (see, e.g., Refs.~\cite{water_emiliomarivi,Zhang2011}). However, {\em ab initio} molecular dynamics (AIMD) studies using PW methods have been limited by the small system sizes and simulations times that are accessible due to computational cost, while the accuracy of existing NAO basis sets for such systems has never been systematically verified, as we propose here.

Using the water monomer as our starting point, we construct double-, triple-, and quadruple-$\zeta$ basis sets for hydrogen and oxygen. We make use of a previously-documented confinement potential for the valence orbitals~\cite{Junquera2001}, and propose a simple new form for the polarization orbitals. We find this new confinement scheme to give a good control over the overall shape of the polarization orbitals by varying a single parameter, making it ideal for variational optimization. We test our basis sets on the water dimer, ice I$_\mathrm{c}$ and VIII, and liquid water, showing our most accurate bases to achieve the level of precision of high-quality PW calculations for energies, pressures, and forces, and all bases to be highly transferable between systems. Therefore, our basis sets can be used with confidence for ambitious future studies of aqueous systems, as they enable cost-efficient large-scale simulations of high accuracy.

\section{Computational methods}

We use the SIESTA~\cite{Soler2002} code for generating the NAO bases and performing the DFT calculations with them, and the ABINIT~\cite{abinit-generic} code for the comparison calculations using PW bases. For both codes we use the PBE~\cite{pbe} semi-local (GGA) exchange-correlation functional, and the same set of norm-conserving pseudopotentials in Troullier-Martins form~\cite{Troullier1991}. The pseudopotential core radius is 0.66~\AA\ for all angular momentum channels of H and 0.61~\AA\ for all channels of O; we also employ a small non-linear core correction~\cite{Louie1982} for O. The pseudopotentials are factorized in the separable Kleinman-Bylander form~\cite{pseudo-KB}, with the same local and non-local components used in both codes~\cite{Verstraete}.

The NAOs making up our basis sets are composed of a freely-varying radial function multiplied by a spherical harmonic; the radial part is defined numerically and is strictly zero beyond a cutoff radius $r_c$. The SIESTA method for generating NAO bases has been documented in several previous publications~\cite{Artacho1999,Junquera2001,Anglada2002}. The main ideas are:
\begin{itemize}
\item a soft confinement potential of the form $V_0 \exp{\{ -(r_c-r_i)/(r-r_i) \}} / (r_c-r)$ between an inner radius $r_i$ and $r_c$~\cite{Junquera2001};
\item a scheme for obtaining multiple-$\zeta$ for each orbital~\cite{Artacho1999}, inspired by the split valence method used in quantum chemistry for Gaussian basis sets~\cite{Huzinaga1984,Poirier1985};
\item polarization orbitals obtained either by applying a small electric field within perturbation theory~\cite{Artacho1999}, or by adding unoccupied atomic shells of higher $l$ with soft confinement~\cite{Junquera2001}.
\end{itemize}
Using these schemes, it is possible to achieve significant optimization of a fixed-size basis by varying the free parameters for each shell: $r_c$, $r_i$ and $V_0$, and the matching radii $\{ r_{\mathrm{d}\zeta}, r_{\mathrm{t}\zeta}, \ldots \}$ which define the multiple-$\zeta$ orbitals. Increasing the orbital cutoff radius generally provides a better quality basis, but also increases its computational cost. Two schemes have been proposed for achieving an optimal balance between accuracy and cost based on a single parameter: the orbital energy shift in the isolated atom caused by confinement~\cite{Artacho1999}, and a fictitious `basis enthalpy' calculated using the orbital volume and a pressure-like variable~\cite{Anglada2002}.

For the valence shells, soft confinement has proven to be very satisfactory: not only does it remove the problematic derivative discontinuity introduced by hard-walled confinement, but it also performs better at a given $r_c$ from a variational point of view. This can be attributed to the fact that it has little effect on the shape of the orbital in the core region, resulting in a good agreement with the free atom orbital.

Instead, the choice of polarization orbitals is more problematic, as the free atom orbitals of higher $l$ can become quite extended, or even entirely unbound. The explicit polarization of the pseudoatom by a small electric field provides an elegant and parameter-free solution, as the extent of the polarization orbitals is defined by that of the orbitals they polarize. Practical applications, however, have shown that a better variational estimate is instead usually obtained by including the unoccupied atomic shells of higher $l$ with an aggressive soft confinement, in order to control the position of the maximum of the radial part of the orbital~\footnote{This shows some similarity to the maximum overlap approach commonly used to define the polarization orbitals in Gaussian basis sets~\cite{Huzinaga1984,Poirier1985}.}. It is also important to note that the former method only allows for orbitals up to $l_v+1$ ($l_v$ being the highest angular momentum for the valence shells), while the latter method can include shells of any $l$.

\begin{figure}
\begin{center}
{\small{\begin{picture}(7200.00,3528.00)%
    \gdef\gplbacktext{}%
    \gdef\gplfronttext{}%
    \gplgaddtomacro\gplbacktext{%
      \csname LTb\endcsname%
      \put(876,705){\makebox(0,0)[r]{\strut{}0}}%
      \put(876,1345){\makebox(0,0)[r]{\strut{}0.4}}%
      \put(876,1985){\makebox(0,0)[r]{\strut{}0.8}}%
      \put(876,2625){\makebox(0,0)[r]{\strut{}1.2}}%
      \put(876,3265){\makebox(0,0)[r]{\strut{}1.6}}%
      \put(1008,485){\makebox(0,0){\strut{}0}}%
      \put(1493,485){\makebox(0,0){\strut{}0.5}}%
      \put(1977,485){\makebox(0,0){\strut{}1}}%
      \put(2462,485){\makebox(0,0){\strut{}1.5}}%
      \put(2947,485){\makebox(0,0){\strut{}2}}%
      \put(3431,485){\makebox(0,0){\strut{}2.5}}%
      \put(238,1985){\rotatebox{-270}{\makebox(0,0){\strut{}$\phi (r)$}}}%
      \put(2462,155){\makebox(0,0){\strut{}$r$ (\AA)}}%
    }%
    \gplgaddtomacro\gplfronttext{%
      \csname LTb\endcsname%
      \put(2929,3092){\makebox(0,0)[r]{\strut{}Soft confinement}}%
      \csname LTb\endcsname%
      \put(2929,2872){\makebox(0,0)[r]{\strut{}Free atom}}%
      \csname LTb\endcsname%
      \put(3189,1345){\makebox(0,0){\strut{}2p}}%
    }%
    \gplgaddtomacro\gplbacktext{%
      \put(3784,705){\makebox(0,0)[r]{\strut{}}}%
      \put(3784,1345){\makebox(0,0)[r]{\strut{}}}%
      \put(3784,1985){\makebox(0,0)[r]{\strut{}}}%
      \put(3784,2625){\makebox(0,0)[r]{\strut{}}}%
      \put(3784,3265){\makebox(0,0)[r]{\strut{}}}%
      \put(3916,485){\makebox(0,0){\strut{}0}}%
      \put(4401,485){\makebox(0,0){\strut{}0.5}}%
      \put(4885,485){\makebox(0,0){\strut{}1}}%
      \put(5370,485){\makebox(0,0){\strut{}1.5}}%
      \put(5855,485){\makebox(0,0){\strut{}2}}%
      \put(6339,485){\makebox(0,0){\strut{}2.5}}%
      \put(6824,485){\makebox(0,0){\strut{}3}}%
      \put(5370,155){\makebox(0,0){\strut{}$r$ (\AA)}}%
    }%
    \gplgaddtomacro\gplfronttext{%
      \csname LTb\endcsname%
      \put(5837,3092){\makebox(0,0)[r]{\strut{}$E$-field polarization}}%
      \csname LTb\endcsname%
      \put(5837,2872){\makebox(0,0)[r]{\strut{}Soft confinement}}%
      \csname LTb\endcsname%
      \put(5837,2652){\makebox(0,0)[r]{\strut{}Coulomb confinement}}%
      \csname LTb\endcsname%
      \put(5837,2432){\makebox(0,0)[r]{\strut{}Free atom}}%
      \csname LTb\endcsname%
      \put(6097,1345){\makebox(0,0){\strut{}3d}}%
    }%
    \gplbacktext
    \put(0,0){\includegraphics{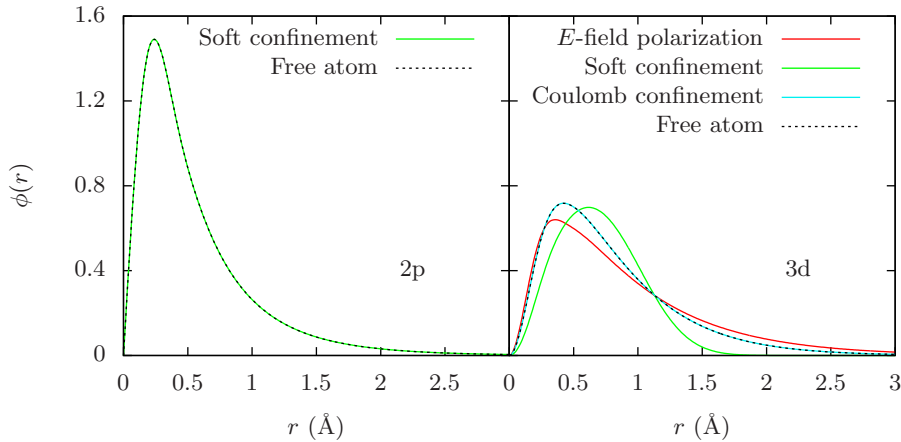}}%
    \gplfronttext
  \end{picture}}}
\caption{Shape of the 2p (occupied) and 3d (unoccupied) orbitals for O. $r_c=5.3$~\AA\ both for the two confinement schemes and the $E$-field polarization orbital.}
\label{fig:Q_conf}
\end{center}
\end{figure}

In this work we propose a new confinement scheme for obtaining short-range polarization orbitals that exhibit a good agreement with the free atom orbitals in the core region. We use a Yukawa-like screened Coulomb potential of the form
\begin{equation}
\label{eq:Q}
V (r) = -Q_0 \frac{\mathrm{e}^{-\lambda r}}{\sqrt{r^2+\delta^2}},
\end{equation}
for which the main parameter for variational optimization is the strength $Q_0$~\footnote{The soft confinement potential is included in addition to the Coulomb potential, but now only for removing the discontinuity at $r_c$ rather than for shaping the orbital. Default values of $V_0=40$~Ry$\cdot$a$_0$ and $r_i=0.9 r_c$ can generally be used without further optimization.}. $\delta$ is introduced to avoid numerical difficulties arising from the singularity at $r=0$, and after some tests has been fixed to 0.01 a$_0$. $\lambda$ can additionally be used for fine tuning of the orbital tail (the default value is set to 0, thus making Eq.~\ref{eq:Q} a normal Coulomb potential).

\begin{figure}
\begin{center}
{\small{\begin{picture}(7200.00,3528.00)%
    \gdef\gplbacktext{}%
    \gdef\gplfronttext{}%
    \gplgaddtomacro\gplbacktext{%
      \csname LTb\endcsname%
      \put(876,705){\makebox(0,0)[r]{\strut{}0}}%
      \put(876,1345){\makebox(0,0)[r]{\strut{}0.2}}%
      \put(876,1985){\makebox(0,0)[r]{\strut{}0.4}}%
      \put(876,2625){\makebox(0,0)[r]{\strut{}0.6}}%
      \put(876,3265){\makebox(0,0)[r]{\strut{}0.8}}%
      \put(1008,485){\makebox(0,0){\strut{}0}}%
      \put(1590,485){\makebox(0,0){\strut{}1}}%
      \put(2171,485){\makebox(0,0){\strut{}2}}%
      \put(2753,485){\makebox(0,0){\strut{}3}}%
      \put(3334,485){\makebox(0,0){\strut{}4}}%
      \put(238,1985){\rotatebox{-270}{\makebox(0,0){\strut{}$\phi (r)$}}}%
      \put(2462,155){\makebox(0,0){\strut{}$r$ (\AA)}}%
    }%
    \gplgaddtomacro\gplfronttext{%
      \csname LTb\endcsname%
      \put(2929,3092){\makebox(0,0)[r]{\strut{}Soft confinement}}%
      \csname LTb\endcsname%
      \put(2929,2872){\makebox(0,0)[r]{\strut{}Free atom}}%
      \csname LTb\endcsname%
      \put(3189,1345){\makebox(0,0){\strut{}3p}}%
    }%
    \gplgaddtomacro\gplbacktext{%
      \put(3784,705){\makebox(0,0)[r]{\strut{}}}%
      \put(3784,1345){\makebox(0,0)[r]{\strut{}}}%
      \put(3784,1985){\makebox(0,0)[r]{\strut{}}}%
      \put(3784,2625){\makebox(0,0)[r]{\strut{}}}%
      \put(3784,3265){\makebox(0,0)[r]{\strut{}}}%
      \put(3916,485){\makebox(0,0){\strut{}0}}%
      \put(4498,485){\makebox(0,0){\strut{}1}}%
      \put(5079,485){\makebox(0,0){\strut{}2}}%
      \put(5661,485){\makebox(0,0){\strut{}3}}%
      \put(6242,485){\makebox(0,0){\strut{}4}}%
      \put(6824,485){\makebox(0,0){\strut{}5}}%
      \put(5370,155){\makebox(0,0){\strut{}$r$ (\AA)}}%
    }%
    \gplgaddtomacro\gplfronttext{%
      \csname LTb\endcsname%
      \put(5837,3092){\makebox(0,0)[r]{\strut{}$E$-field polarization}}%
      \csname LTb\endcsname%
      \put(5837,2872){\makebox(0,0)[r]{\strut{}Soft confinement}}%
      \csname LTb\endcsname%
      \put(5837,2652){\makebox(0,0)[r]{\strut{}Coulomb confinement}}%
      \csname LTb\endcsname%
      \put(5837,2432){\makebox(0,0)[r]{\strut{}Free atom (unbound)}}%
      \csname LTb\endcsname%
      \put(6097,1345){\makebox(0,0){\strut{}3d}}%
    }%
    \gplbacktext
    \put(0,0){\includegraphics{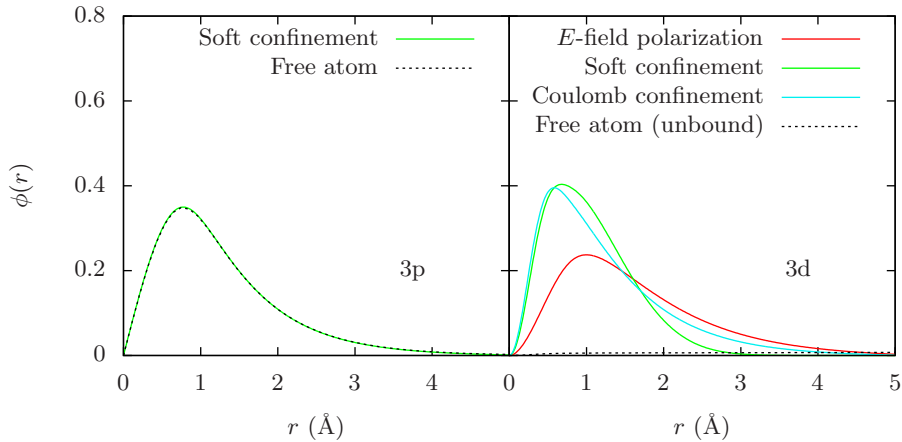}}%
    \gplfronttext
  \end{picture}}}
\caption{Shape of the 3p (occupied) and 3d (unoccupied) orbitals for Si. $r_c=5.3$~\AA\ both for the two confinement schemes and the $E$-field polarization orbital.}
\label{fig:Q_confb}
\end{center}
\end{figure}

Fig.~\ref{fig:Q_conf} shows the radial part of the orbitals obtained with our new Coulomb confinement scheme for the 3d shell of O. Its decay is intermediate between those obtained by $E$-field polarization and soft confinement, and is similar to that of the free atom orbital. Within a double-$\zeta$ basis with a single polarization shell (\DZP), the Coulomb confinement scheme results in a variationally better basis for the water monomer and dimer respect to soft confinement (by $\sim$80~meV/molecule), as well as a smaller counterpoise (CP) correction~\cite{cc-bsse,cc-bsse2} (by 5~meV). Fig.~\ref{fig:Q_confb} shows a similar behaviour for the 3d shell of Si; in this case, however, the free atom orbital is unbound. As before, the radial part of the Coulomb-confined d orbitals exhibits a slower decay than the soft-confined one, thereby achieving a greater overlap with that of the p orbitals they are polarizing, which gives the latter more flexibility. In bulk silicon, this results in Coulomb confinement gaining $\sim$4~meV/atom respect to soft confinement.

Making use of the Coulomb confinement scheme for the polarization orbitals, we have developed a series of 13 basis sets of increasing accuracy for the water molecule, ranging from 23 to 91 basis orbitals/molecule. Following the systematic convergence strategy proposed for correlation-consistent basis sets~\cite{Dunning1989}, the maximum size of our bases depends on the number of $\zeta$ orbitals used for the valence shells, decreasing by one for each additional polarization shell: \DZP, \TZDPP, \QZTPDPP, and so on. We also provide intermediary bases with various subsets of the full number of polarization orbitals indicated by this scheme. We limit $l \le 3$, and so the largest basis we consider is \QZTPDP. For cases in which the number of $\zeta$ orbitals in the polarization shells differs between O and H, we adopt the following naming convention: the highest level of $\zeta$ is used, followed in brackets by the element for which it applies; the other element in such cases has one fewer shell. For example, \TZDPH\ includes a double-$\zeta$ polarization shell for H, and a single-$\zeta$ polarization shell for O.

We do not here consider the optimization of $r_c$, but simply fix it to 4.5~\AA\ for the double- and triple-$\zeta$ bases, and to 5.3~\AA\ for the quadruple-$\zeta$ bases. We test the effect of varying the cutoff radius for the \QZDP\ basis, by decreasing $r_c$ to 4.5~\AA\ (the shorter basis is denoted as \QZDPs); the difference in energy is minimal (Table~\ref{table:dimer_summary}). Other parameters, including $Q_0$ and $\lambda$, are optimized variationally for the water monomer (see Supplementary Material for full specifications of all basis sets used). We also include in our results two previously proposed bases~\cite{water_emiliomarivi}, obtained using the basis enthalpy optimization procedure~\cite{Anglada2002} with a confining pressure of 0.2~GPa; for these bases (which we denote as \DZPmv\ and \TZPmv) the value of $r_c$ varies between shells, up to a maximum of 3.3~\AA\ for \DZPmv\ and 3.7~\AA\ for \TZPmv.

\subsection{General guidelines for generating a basis set}

\begin{figure}
\begin{center}
{\small{\begin{picture}(5040.00,3528.00)%
    \gdef\gplbacktext{}%
    \gdef\gplfronttext{}%
    \gplgaddtomacro\gplbacktext{%
      \csname LTb\endcsname%
      \put(1078,704){\makebox(0,0)[r]{\strut{} 0.01}}%
      \put(1078,1344){\makebox(0,0)[r]{\strut{} 0.1}}%
      \put(1078,1984){\makebox(0,0)[r]{\strut{} 1}}%
      \put(1078,2623){\makebox(0,0)[r]{\strut{} 10}}%
      \put(1078,3263){\makebox(0,0)[r]{\strut{} 100}}%
      \put(1210,484){\makebox(0,0){\strut{} 0}}%
      \put(2068,484){\makebox(0,0){\strut{} 500}}%
      \put(2927,484){\makebox(0,0){\strut{} 1000}}%
      \put(3785,484){\makebox(0,0){\strut{} 1500}}%
      \put(4643,484){\makebox(0,0){\strut{} 2000}}%
      \put(176,1983){\rotatebox{-270}{\makebox(0,0){\strut{}Monomer energy error (eV)}}}%
      \put(2926,154){\makebox(0,0){\strut{}PW cutoff (eV)}}%
    }%
    \gplgaddtomacro\gplfronttext{%
      \csname LTb\endcsname%
      \put(3656,3090){\makebox(0,0)[r]{\strut{}$\mathrm{d}\zeta$}}%
      \csname LTb\endcsname%
      \put(3656,2870){\makebox(0,0)[r]{\strut{}$\mathrm{t}\zeta$}}%
      \csname LTb\endcsname%
      \put(3656,2650){\makebox(0,0)[r]{\strut{}$\mathrm{q}\zeta$}}%
    }%
    \gplbacktext
    \put(0,0){\includegraphics{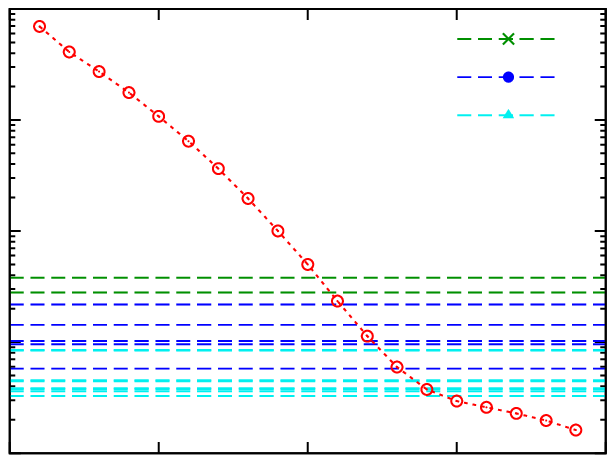}}%
    \gplfronttext
  \end{picture}}{\begin{picture}(2520.00,3528.00)%
    \gdef\gplbacktext{}%
    \gdef\gplfronttext{}%
    \gplgaddtomacro\gplbacktext{%
      \csname LTb\endcsname%
      \put(858,704){\makebox(0,0)[r]{\strut{} 0.01}}%
      \put(858,1785){\makebox(0,0)[r]{\strut{} 0.1}}%
      \put(858,2867){\makebox(0,0)[r]{\strut{} 1}}%
      \put(990,484){\makebox(0,0){\strut{}70}}%
      \put(1557,484){\makebox(0,0){\strut{}100}}%
      \put(2123,484){\makebox(0,0){\strut{}130}}%
      \put(990,3087){\makebox(0,0){\strut{} 0}}%
      \put(1557,3087){\makebox(0,0){\strut{} 50}}%
      \put(2123,3087){\makebox(0,0){\strut{} 100}}%
      \put(176,1785){\rotatebox{-270}{\makebox(0,0){\strut{}}}}%
      \put(1556,154){\makebox(0,0){\strut{}PW basis size/V (\AA$^{-3}$)}}%
      \put(1556,3416){\makebox(0,0){\strut{}NAO basis size}}%
    }%
    \gplgaddtomacro\gplfronttext{%
    }%
    \gplbacktext
    \put(0,0){\includegraphics{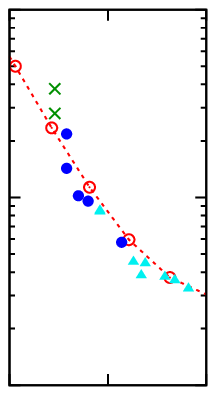}}%
    \gplfronttext
  \end{picture}}}
\caption{Convergence of the total energy for the monomer. The points indicate ABINIT PW calculations at different cutoffs, and the dashed horizontal lines show the values obtained with all SIESTA NAO basis sets. On the right we show the same results in a smaller range, with the SIESTA results now plotted against basis size, and the ABINIT results plotted against basis size per unit volume. All energy values are referenced to a PW calculation with extremely high cutoff (5000~eV).}
\label{fig:monomer}
\end{center}
\end{figure}

The many different options for defining the basis orbitals discussed and reviewed above are at the disposal of the user for generating good basis sets. It must be remembered, however, that there is no unique way of defining a basis, and that, as for pseudopotentials, the responsibility of the choice of basis lies with the person performing the calculations. Methods like SIESTA are limited to using NAOs of finite range, but there is absolute freedom in the radial shapes, their range, the number of orbitals, the angular momentum values, and even the location (which is not restricted to be centred on an atom).

Here we present some guidelines for producing reasonable basis sets:
\begin{itemize}
\item Use as many basis functions as required for the accuracy needed, following the canonical hierarchy described above starting from the minimal single-$\zeta$ basis: \SZ, \DZP, \TZDPP, \QZTPDPP, and so on.
\item Avoid too large or too small orbital ranges. Some procedures, like the one based on an energy shift~\cite{Artacho1999}, or blind variational optimization based on a single reference~\cite{Anglada2002} can produce very short orbitals for light elements, inner orbitals, or cations. This severely limits the transferability of the basis. Never use orbitals with radii smaller than 3~\AA. On the other hand, some procedures generate extremely large cutoff radii (e.g., in loosely bound states, as the 4s shell in transition metals). If they are used for bulk calculations, functions with radii larger than 5~\AA\ can be quite useless (there are sufficient other functions in neighbouring atoms) but make the calculations substantially less efficient. SIESTA allows the user to establish both a minimum and a maximum cutoff radius globally.
\item Use the soft confinement potential for smoothening the radial function close to $r_c$ only, with $r_i \approx 0.9 r_c$.
\item For generating a polarization shell, use the Coulomb confinement potential described above (if not using the perturbative $E$-field polarization option). $Q_0$ (and, optionally, $\lambda$) can be defined variationally on a set of representative simple systems, or by maximizing the overlap of the radial part of the polarization orbitals $\phi_{l'} (r)$ with the one of the shell to be polarized $\phi_l (r)$: $\int_0^{r_c} r^2 \phi_{l'} (r) \phi_l (r) \mathrm{d}r$.
\item Finally, the matching radii for the multiple-$\zeta$ orbitals can be defined variationally.
\end{itemize}

\section{Results}

\subsection{Water monomer and dimer}

\begin{figure}
\begin{center}
{\small{\begin{picture}(5040.00,3528.00)%
    \gdef\gplbacktext{}%
    \gdef\gplfronttext{}%
    \gplgaddtomacro\gplbacktext{%
      \csname LTb\endcsname%
      \put(946,704){\makebox(0,0)[r]{\strut{} 80}}%
      \put(946,1216){\makebox(0,0)[r]{\strut{} 120}}%
      \put(946,1728){\makebox(0,0)[r]{\strut{} 160}}%
      \put(946,2239){\makebox(0,0)[r]{\strut{} 200}}%
      \put(946,2751){\makebox(0,0)[r]{\strut{} 240}}%
      \put(946,3263){\makebox(0,0)[r]{\strut{} 280}}%
      \put(1078,484){\makebox(0,0){\strut{} 0}}%
      \put(1587,484){\makebox(0,0){\strut{} 200}}%
      \put(2097,484){\makebox(0,0){\strut{} 400}}%
      \put(2606,484){\makebox(0,0){\strut{} 600}}%
      \put(3115,484){\makebox(0,0){\strut{} 800}}%
      \put(3624,484){\makebox(0,0){\strut{} 1000}}%
      \put(4134,484){\makebox(0,0){\strut{} 1200}}%
      \put(4643,484){\makebox(0,0){\strut{} 1400}}%
      \put(176,1983){\rotatebox{-270}{\makebox(0,0){\strut{}Dimer binding energy (meV)}}}%
      \put(2860,154){\makebox(0,0){\strut{}PW cutoff (eV)}}%
      \put(2861,3001){\makebox(0,0)[r]{\strut{}$\mathrm{d}\zeta+\mathrm{p}$}}%
      \put(2097,2871){\makebox(0,0)[r]{\strut{}$\mathrm{t}\zeta+\mathrm{p}$}}%
      \put(2097,2537){\makebox(0,0)[r]{\strut{}$\mathrm{q}\zeta+\mathrm{dp}$}}%
    }%
    \gplgaddtomacro\gplfronttext{%
    }%
    \gplbacktext
    \put(0,0){\includegraphics{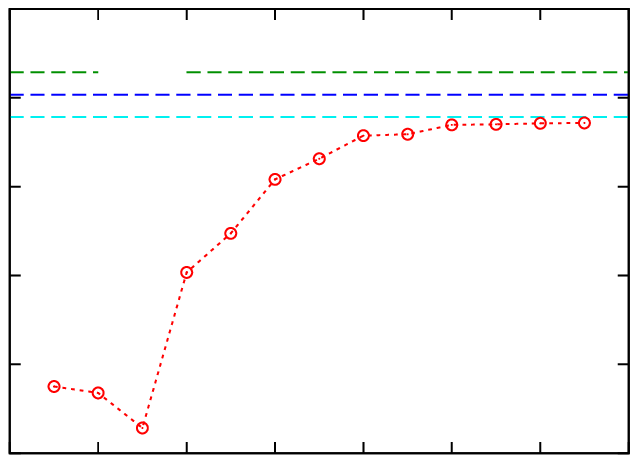}}%
    \gplfronttext
  \end{picture}}{\begin{picture}(2520.00,3528.00)%
    \gdef\gplbacktext{}%
    \gdef\gplfronttext{}%
    \gplgaddtomacro\gplbacktext{%
      \csname LTb\endcsname%
      \put(726,1024){\makebox(0,0)[r]{\strut{} 224}}%
      \put(726,1664){\makebox(0,0)[r]{\strut{} 226}}%
      \put(726,2303){\makebox(0,0)[r]{\strut{} 228}}%
      \put(726,2943){\makebox(0,0)[r]{\strut{} 230}}%
      \put(1174,484){\makebox(0,0){\strut{} 1000}}%
      \put(1807,484){\makebox(0,0){\strut{} 1200}}%
      \put(176,1983){\rotatebox{-270}{\makebox(0,0){\strut{}}}}%
      \put(1490,154){\makebox(0,0){\strut{}}}%
      \put(1965,1733){\makebox(0,0)[r]{\strut{}$\mathrm{d}\zeta+\mathrm{p}$}}%
      \put(1965,2682){\makebox(0,0)[r]{\strut{}$\mathrm{t}\zeta+\mathrm{p}$}}%
      \put(1649,2896){\makebox(0,0)[r]{\strut{}$\mathrm{q}\zeta+\mathrm{dp}$}}%
      \put(1807,928){\makebox(0,0){\strut{}(CP)}}%
    }%
    \gplgaddtomacro\gplfronttext{%
    }%
    \gplbacktext
    \put(0,0){\includegraphics{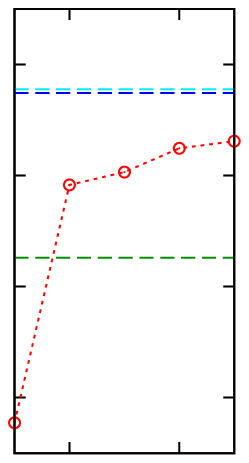}}%
    \gplfronttext
  \end{picture}}}
\caption{Convergence of the dimer binding energy. The points indicate ABINIT PW calculations at different cutoffs, and the dashed horizontal lines show the values obtained with three SIESTA NAO basis sets. On the right we show the same results in a smaller range, with the SIESTA results now including the counterpoise correction.}
\label{fig:dimer}
\end{center}
\end{figure}

We first show the convergence of the total energy for a single water molecule in a large box ($10 \times 10 \times 10$~\AA$^3$). We perform a series of PW calculations of increasing kinetic energy cutoff, up to a maximum of 5000~eV (by which point the total energy is converged to $<1$~meV). In Fig~\ref{fig:monomer}, we show how our NAO bases compare with the PW bases. Since the same pseudopotential is used in both sets of calculations, the total energies can be compared directly. The NAO bases show a smooth convergence with basis size, similar to that of PWs; it is tempting therefore to equate each NAO basis with the PW cutoff giving the same total energy, despite the fact that the Hilbert space spanned by them is very different. Nevertheless, the NAO bases undoubtably achieve a good level of variational convergence, with the \QZTPDP\ basis reducing the error in the total energy to $\sim$30~meV, as much as a 1460~eV PW cutoff. Even our smallest basis, \DZP, is reasonably well converged: the total energy error is $\sim$380~meV, equivalent to a 1080~eV cutoff. The error for all NAO basis sets is listed in Table~\ref{table:dimer_summary}.

There is a fairly large energy difference between the \DZPmv\ and \TZPmv\ bases and our newly parametrized ones of the same size (denoted simply \DZP\ and \TZP). This we attribute both to the use of Coulomb confinement for the first polarization shell and the larger cutoff radii. In general, the convergence with basis size is almost monotomic; the only exception is the \QZDPPDO\ basis, for which we add a set of diffuse orbitals of s and p symmetry onto the O ion in order to capture the swelling due to its anionic character. This basis adds four new orbitals onto \QZDPP, yet gains 6~meV more energy than \QZTPHP, which adds six. Such small energy differences, however, do not significantly alter the overall convergence behaviour.

Fig.~\ref{fig:dimer} shows the PW convergence and values for three selected NAO bases of the binding energy $E_b$ of the water dimer in a fixed geometry (from Supplementary Material of Ref.~\cite{Santra2007}). The results for all NAO bases are given in Table~\ref{table:dimer_summary}. In order to ensure that periodic image interactions are negligible, we use a $25 \times 25 \times 25$~\AA\ box. Here, the difference between NAO and PW bases is clear, as the former converges from above and the latter from below. All our quadruple-$\zeta$ bases give an error in $E_b$ within 2--3~meV of the converged PW result. For the double- and triple-$\zeta$ bases, basis set superposition error (BSSE) accounts for most of the total error. The addition of a CP correction term to $E_b$ brings all bases to within the same level of precision of the quadruple-$\zeta$ ones; only \DZPmv\ and \TZPmv\ are somewhat less precise (with errors of 10--13~meV), despite having a smaller CP correction due to their shorter range.

\begin{table}
\caption{Summary of results for the monomer and dimer. Listed are the basis sets tested, the basis size (number of orbitals/molecule), the total energy of the monomer ($E$, referenced to a PW calculation with 5000~eV cutoff), the binding energy of the dimer, first without and then with the counterpoise correction ($E_b$ and $E_b^\mathrm{CP}$, respectively), and the value of the correction. The PW binding energy listed is calculated with 1300~eV cutoff.}
\begin{center}
{\footnotesize
\lineup
\begin{tabular}{@{}lccccc}
\br
Basis set & Basis size & $E$ (meV) & $E_b$ (meV) & $E_b^\mathrm{CP}$ (meV) & CP corr. (meV) \\
\mr                                              
\DZPmv    &         23 &   378.61  & 266.38      & 241.16                  &  $-$25.22      \\
\DZP      &         23 &   280.02  & 251.51      & 226.52                  &  $-$24.99      \\
\TZPmv    &         29 &   217.98  & 249.42      & 239.89                  & \0$-$9.53      \\
\TZP      &         29 &   142.84  & 241.34      & 229.48                  &  $-$11.86      \\
\TZDPH    &         35 &   102.14  & 232.01      & 226.80                  & \0$-$5.21      \\
\TZDP     &         40 &  \095.59  & 230.68      & 225.96                  & \0$-$4.72      \\
\TZDPP    &         57 &  \057.64  & 228.68      & 223.65                  & \0$-$5.03      \\
\QZDPs    &         46 &  \084.97  & 231.50      & 229.70                  & \0$-$1.80      \\
\QZDP     &         46 &  \083.98  & 231.28      & 229.56                  & \0$-$1.72      \\
\QZDPP    &         63 &  \045.37  & 227.99      & 226.25                  & \0$-$1.74      \\
\QZDPPDO  &         67 &  \038.43  & 229.26      & 227.16                  & \0$-$2.10      \\
\QZTPHP   &         69 &  \044.43  & 227.68      & 226.58                  & \0$-$1.10      \\
\QZTPHDPH &         79 &  \037.71  & 228.01      & 226.54                  & \0$-$1.47      \\
\QZTPDPH  &         84 &  \036.17  & 228.10      & 226.68                  & \0$-$1.42      \\
\QZTPDP   &         91 &  \032.64  & 228.63      & 227.12                  & \0$-$1.51      \\
PW        &        \0- & \0\00.00  & 228.62      & \0\0\m-\0\0             & \0\0\m-\0\0    \\
\br
\end{tabular}
\label{table:dimer_summary}
}
\end{center}
\end{table}

\subsection{Ice phases: I$_c$ and VIII}

We now turn to the first test of our bases in bulk water systems, by focussing on two of the ordered phases of ice: cubic ice I$_\mathrm{c}$ and high-density tetragonal ice VIII. The study of ice by first principles DFT methods has progressed considerably in recent years~\cite{Militzer2010,Hamada2010,Hermann2011,Santra2011,Murray2012}, but understanding in detail the relative energetic contributions that give rise to its phase diagram and the properties of the various phases remains a challenging problem of current interest. Typically for such studies, it is desirable to calculate energy differences between configurations to within a few meV/molecule. To this end, we test the accuracy of the NAO bases both in terms of the relaxation of the ionic positions and the final energy difference between the two phases under consideration.

\begin{table}
\caption{Summary of results for ice. Listed are total energies/molecule (meV) for the two phases considered and the energy difference/molecule (meV) between these phases for seven different geometries, obtained by relaxing with six NAO basis sets, and a PW relaxation with 4500~eV cutoff. For each geometry we give the energy calculated with SIESTA using the same NAO basis for which the relaxation was performed, and with ABINIT at 4500~eV cutoff (independently of geometry). Note that the total energies of the two phases are given referenced to their respective PW results.}
\begin{center}
{\footnotesize
\lineup
\begin{tabular}{@{}lcccccccc}
\br
                                &                        & \multicolumn{7}{c}{Geometry} \\ \noalign{\smallskip}\cline{3-9}\noalign{\smallskip}
Phase                           & Basis                  & \DZPmv        & \DZP       & \TZPmv        & \TZP       & \QZDP         & \QZTPDP       & PW                           \\
\mr                                                                                                                                
\multirow{2}{*}{I$_\mathrm{c}$} & NAO                    & \m295.93      & \m197.35   & \m148.34      & \m107.68   & \m\068.70     & \m\025.29     & \m\0\0\0-\0\0                \\
                                & PW                     & \m\0\00.78    & \m\0\02.73 & \m\0\02.11    & \m\0\00.99 & \m\0\00.55    & \m\0\00.11    & \m\0\00.00                   \\ \noalign{\smallskip}\cline{2-9}\noalign{\smallskip}
\multirow{2}{*}{VIII}           & NAO                    & \m259.76      & \m166.58   & \m161.32      & \m111.92   & \m\073.89     & \m\026.86     & \m\0\0\0-\0\0                \\
                                & PW                     & \m\0\01.49    & \m\0\02.88 & \m\0\01.95    & \m\0\01.04 & \m\0\00.53    & \m\0\00.17    & \m\0\00.00                   \\
\mr                                                                                                                                
\multirow{3}{*}{Diff.}          & NAO                    & $-$142.58     & $-$147.99  & $-$191.73     & $-$182.99  & $-$183.95     & $-$180.32     & \m\0\0\0-\0\0                \\
                                & PW                     & $-$179.46     & $-$178.90  & $-$178.59     & $-$178.80  & $-$178.73     & $-$178.82     & $-$178.75                    \\
                                & Ref.~\cite{Murray2012} & \m\0\0\0-\0\0 & -          & \m\0\0\0-\0\0 & -          & \m\0\0\0-\0\0 & \m\0\0\0-\0\0 & $-$178{\color{white} .}\0\0  \\
\br
\end{tabular}
\label{table:ice_summary}
}
\end{center}
\end{table}

For the calculations, we fix the volume/molecule to 30.51~\AA$^3$ for ice Ic and 20.45~\AA$^3$ for ice VIII, using the equilibrium volumes reported by Murray and Galli~\cite{Murray2012} for PBE at ambient pressure, and the c/a ratio to 1.44 for ice VIII. We use a $4 \times 4 \times 4$ Monkhorst-Pack (MP) k-point grid~\cite{mp_grid} for both unit cells. First, we perform PW cutoff convergence tests for the two ice phases up to a 5000~eV cutoff; we find that 4500~eV gives extremely accurate results (within 0.1~meV for total energies, 0.5~meV/\AA\ for ionic forces, and 0.05~kbar for pressures), and use this value for all PW results given in this section.

We choose four representative NAO basis sets for testing from those presented in the previous section; additionally, the two old parametrizations for double- and triple-$\zeta$ are also tested. The ice unit cells are relaxed with respect to the ionic positions for these six bases, and independently for PWs, using a maximum force tolerance of 10~meV/\AA. Finally, we recalculate the energy of the system with PWs using the six geometries obtained from the NAO bases. All results are given in Table~\ref{table:ice_summary}.

The PW results for the different geometries show that all NAO bases give sufficiently accurate ionic relaxations. The errors in the PW energy difference/molecule between the two phases using the NAO-relaxed geometries, as compared with the PW-relaxed ones, are less than a meV in all cases, ranging from 0.71~meV for the \DZPmv\ geometries to only 0.02~meV for the \QZDP\ ones. The errors in total energies are similarly small, less than 3~meV for all geometries. The accuracy of the relaxations is also confirmed by examining the bond lengths, with the largest errors found for the \DZPmv\ geometries (up to 21~m\AA), and the smallest for the \QZTPDP\ ones (up to 3~m\AA).

We now consider not only the geometries, but also the energies calculated with the NAO bases. As should be expected, total energies/molecule of the ice phases have errors on the order of those reported in Table~\ref{table:dimer_summary} for the total energy of the monomer; however, it is interesting to note that they are systematically smaller, resulting from the fact that we are now considering a condensed phase, thus enabling the NAOs to represent the charge density everywhere in the unit cell.

\begin{figure}
\begin{center}
{\small{\begin{picture}(5040.00,3528.00)%
    \gdef\gplbacktext{}%
    \gdef\gplfronttext{}%
    \gplgaddtomacro\gplbacktext{%
      \csname LTb\endcsname%
      \put(1078,704){\makebox(0,0)[r]{\strut{} 0.01}}%
      \put(1078,1344){\makebox(0,0)[r]{\strut{} 0.1}}%
      \put(1078,1984){\makebox(0,0)[r]{\strut{} 1}}%
      \put(1078,2623){\makebox(0,0)[r]{\strut{} 10}}%
      \put(1078,3263){\makebox(0,0)[r]{\strut{} 100}}%
      \put(1210,484){\makebox(0,0){\strut{} 0}}%
      \put(1897,484){\makebox(0,0){\strut{} 400}}%
      \put(2583,484){\makebox(0,0){\strut{} 800}}%
      \put(3270,484){\makebox(0,0){\strut{} 1200}}%
      \put(3956,484){\makebox(0,0){\strut{} 1600}}%
      \put(4643,484){\makebox(0,0){\strut{} 2000}}%
      \put(176,1983){\rotatebox{-270}{\makebox(0,0){\strut{}RMS error/molecule (meV)}}}%
      \put(2926,154){\makebox(0,0){\strut{}PW cutoff (eV)}}%
      \put(1982,2258){\makebox(0,0)[r]{\strut{}$\mathrm{d}\zeta+\mathrm{p}$}}%
      \put(1982,2064){\makebox(0,0)[r]{\strut{}$\mathrm{t}\zeta+\mathrm{p}$}}%
      \put(1982,1649){\makebox(0,0)[r]{\strut{}$\mathrm{q}\zeta+\mathrm{dp}$}}%
    }%
    \gplgaddtomacro\gplfronttext{%
      \csname LTb\endcsname%
      \put(3656,3090){\makebox(0,0)[r]{\strut{}1.20 $\mathrm{g}/\mathrm{cm}^3$}}%
      \csname LTb\endcsname%
      \put(3656,2870){\makebox(0,0)[r]{\strut{}1.00 $\mathrm{g}/\mathrm{cm}^3$}}%
    }%
    \gplbacktext
    \put(0,0){\includegraphics{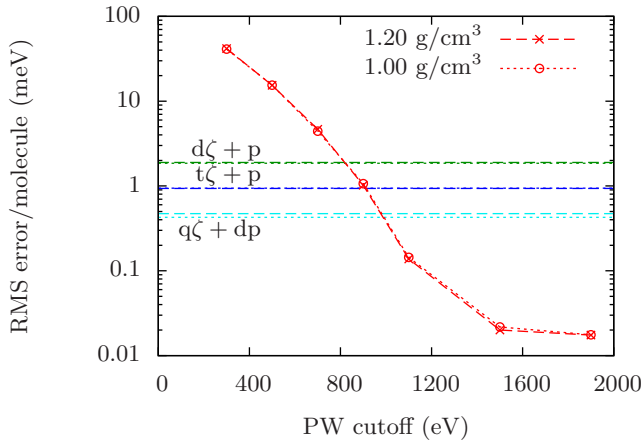}}%
    \gplfronttext
  \end{picture}}}
\caption{RMS error in energy differences/molecule for 100 snapshots at each density. The points indicate PW calculations at different cutoffs, and the dashed horizontal lines show the value for three NAO basis sets (long dashes for 1.20~g/cm$^3$ density and short dashes for 1.00~g/cm$^3$ density).}
\label{fig:Ed_conv}
\end{center}
\end{figure}

The most important results are for the energy differences/molecule between phases given by the NAO calculations. \DZP\ gives a fairly substantial error ($\sim$30~meV), while both \TZP\ and \QZDP\ are in good agreement with the converged PW result, with errors of $\sim$5~meV. The most accurate basis, \QZTPDP, succeeds in reducing the error to less than 2~meV. As with the dimer binding energy, we find our new bases to give a substantial improvement with respect to the old parametrizations, especially in this case for \TZP.

\subsection{Liquid water}

\begin{figure}
\begin{center}
{\small{\begin{picture}(5040.00,3528.00)%
    \gdef\gplbacktext{}%
    \gdef\gplfronttext{}%
    \gplgaddtomacro\gplbacktext{%
      \csname LTb\endcsname%
      \put(814,704){\makebox(0,0)[r]{\strut{}-60}}%
      \put(814,1131){\makebox(0,0)[r]{\strut{}-40}}%
      \put(814,1557){\makebox(0,0)[r]{\strut{}-20}}%
      \put(814,1984){\makebox(0,0)[r]{\strut{} 0}}%
      \put(814,2410){\makebox(0,0)[r]{\strut{} 20}}%
      \put(814,2837){\makebox(0,0)[r]{\strut{} 40}}%
      \put(814,3263){\makebox(0,0)[r]{\strut{} 60}}%
      \put(946,484){\makebox(0,0){\strut{}-60}}%
      \put(1562,484){\makebox(0,0){\strut{}-40}}%
      \put(2178,484){\makebox(0,0){\strut{}-20}}%
      \put(2795,484){\makebox(0,0){\strut{} 0}}%
      \put(3411,484){\makebox(0,0){\strut{} 20}}%
      \put(4027,484){\makebox(0,0){\strut{} 40}}%
      \put(4643,484){\makebox(0,0){\strut{} 60}}%
      \csname LTb\endcsname%
      \put(176,1983){\rotatebox{-270}{\makebox(0,0){\strut{}$\Delta E$/molecule (meV)}}}%
      \put(2794,154){\makebox(0,0){\strut{}$\Delta E_0$/molecule (meV)}}%
    }%
    \gplgaddtomacro\gplfronttext{%
      \csname LTb\endcsname%
      \put(2002,3090){\makebox(0,0)[r]{\strut{}$\mathrm{d}\zeta+\mathrm{p}$}}%
      \csname LTb\endcsname%
      \put(2002,2870){\makebox(0,0)[r]{\strut{}$\mathrm{t}\zeta+\mathrm{p}$}}%
      \csname LTb\endcsname%
      \put(2002,2650){\makebox(0,0)[r]{\strut{}$\mathrm{q}\zeta+\mathrm{dp}$}}%
    }%
    \gplbacktext
    \put(0,0){\includegraphics{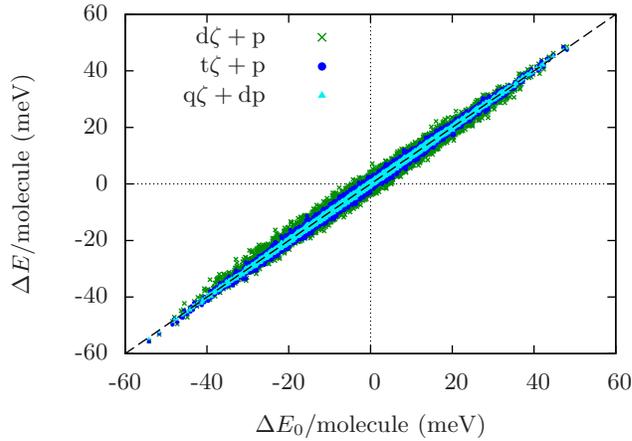}}%
    \gplfronttext
  \end{picture}}}
\caption{Scatter plot comparing energy differences/molecule between all snapshots obtained with ABINIT PW calculations at 2700~eV cutoff ($\Delta E_0$) and SIESTA calculations with three different NAO basis sets.}
\label{fig:Ed_scatter}
\end{center}
\end{figure}

\begin{figure}
\begin{center}
{\small{\begin{picture}(6840.00,3528.00)%
    \gdef\gplbacktext{}%
    \gdef\gplfronttext{}%
    \gplgaddtomacro\gplbacktext{%
      \csname LTb\endcsname%
      \put(814,704){\makebox(0,0)[r]{\strut{}-60}}%
      \put(814,1131){\makebox(0,0)[r]{\strut{}-40}}%
      \put(814,1557){\makebox(0,0)[r]{\strut{}-20}}%
      \put(814,1984){\makebox(0,0)[r]{\strut{} 0}}%
      \put(814,2410){\makebox(0,0)[r]{\strut{} 20}}%
      \put(814,2837){\makebox(0,0)[r]{\strut{} 40}}%
      \put(814,3263){\makebox(0,0)[r]{\strut{} 60}}%
      \put(946,484){\makebox(0,0){\strut{}-60}}%
      \put(1566,484){\makebox(0,0){\strut{}-40}}%
      \put(2185,484){\makebox(0,0){\strut{}-20}}%
      \put(2805,484){\makebox(0,0){\strut{} 0}}%
      \put(3425,484){\makebox(0,0){\strut{} 20}}%
      \put(4044,484){\makebox(0,0){\strut{} 40}}%
      \put(4664,484){\makebox(0,0){\strut{} 60}}%
      \csname LTb\endcsname%
      \put(176,1983){\rotatebox{-270}{\makebox(0,0){\strut{}$\Delta E$/molecule (meV)}}}%
      \put(2805,154){\makebox(0,0){\strut{}$\Delta E_0$/molecule (meV)}}%
    }%
    \gplgaddtomacro\gplfronttext{%
      \csname LTb\endcsname%
      \put(5852,3153){\makebox(0,0)[r]{\strut{}300 eV}}%
      \csname LTb\endcsname%
      \put(5852,2933){\makebox(0,0)[r]{\strut{}500 eV}}%
      \csname LTb\endcsname%
      \put(5852,2713){\makebox(0,0)[r]{\strut{}700 eV}}%
      \csname LTb\endcsname%
      \put(5852,2493){\makebox(0,0)[r]{\strut{}900 eV}}%
      \csname LTb\endcsname%
      \put(5852,2273){\makebox(0,0)[r]{\strut{}1100 eV}}%
      \csname LTb\endcsname%
      \put(5852,2053){\makebox(0,0)[r]{\strut{}1500 eV}}%
    }%
    \gplbacktext
    \put(0,0){\includegraphics{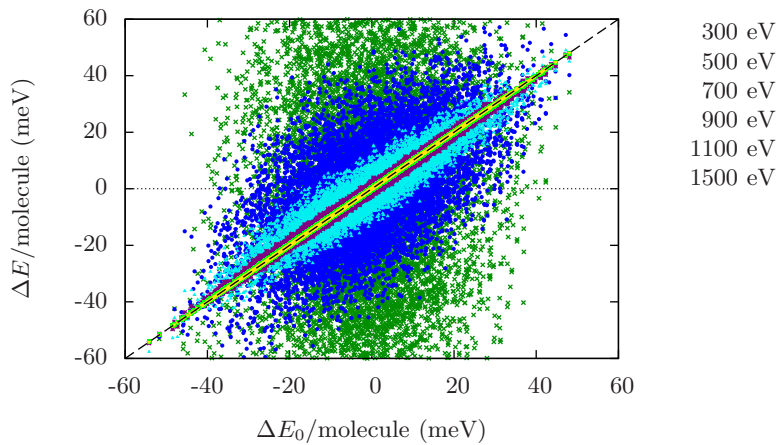}}%
    \gplfronttext
  \end{picture}}}
\caption{Scatter plot comparing energy differences/molecule between all snapshots obtained with ABINIT PW calculations at 2700~eV cutoff ($\Delta E_0$) and lower cutoffs.}
\label{fig:Ed2_scatter}
\end{center}
\end{figure}

Assessing the accuracy of our bases for water in its liquid state is more challenging than for the solid state, as the calculation of any quantity of interest will involve a statistical average over a MD trajectory. Performing a full AIMD simulation using a tightly converged PW cutoff would be computationally very expensive, and, furthermore, would not allow for a detailed comparison with the results obtained for the same simulation using one of the NAO basis sets, as the trajectories would quickly decorrelate.

Here we employ a simple alternative method for testing basis sets for liquid systems, which can be used routinely to obtain accurate estimates of the errors in quantites such as total energies, energy differences, ionic forces and cell pressures, as well as for direct parametrization of the basis set using the same procedures proposed for fitting classical force fields from {\em ab initio} data~\cite{paul-fit}. We first employ the GROMACS~\cite{gromacs} code to perform two long 1~ns MD runs of a box of 32 water molecules using the TIP4P force field~\cite{Jorgensen1983b}, at equilibrium density (1.00~g/cm$^3$) and high density (1.20~g/cm$^3$). We select 100 random snapshots from each run, which we use as our testing set; the long MD run ensures that they are uncorrelated. We include the high density run to ensure that we are sampling sufficient configurations with occupied interstitial anti-tetrahedral sites between the first and second coordination shells, as the correct description of these configurations is important for reproducing several key properties of the liquid~\cite{water_emiliomarivi}. Finally, we perform single-point DFT calculations of the 200 snapshots, using the various NAO bases in SIESTA and PWs at different cutoffs in ABINIT. In this case we choose 2700~eV as our `converged' PW cutoff value for comparing to all other bases. Cutoff convergence tests on a few snapshots up to 5000~eV show this value to give total energies to within 1~meV, ionic forces to within 0.6~meV/\AA, and pressures to within 0.5~kbar.

We perform our tests on five NAO bases, \DZPmv, \DZP, \TZPmv, \TZP, and \QZDP; all results are given in Table~\ref{table:water_summary}. Firstly, we examine the error in total energies/molecule for all the snapshots. As expected, calculating the average shift in total energy for all snapshots gives a similar convergence to that discussed previously for the water monomer and the two ice phases. One point of interest is that the average shifts calculated for the two densities become closer to each other as the quality of the basis increases, anticipating the fact that pressure values also become more precise.

The best quantitative estimator of overall accuracy for a fixed-volume NVE AIMD simulation is the energy difference between snapshots. For this, we calculate the root mean square (RMS) error in our test set; the results are shown in Fig.~\ref{fig:Ed_conv}, compared to PW calculations of increasing cutoff. There is a negligible difference between the two densities, with the \DZP, \TZP, and \QZDP\ bases giving RMS errors equivalent to PW cutoffs of 820~eV, 910~eV, and 980~eV, respectively. This is similar to the errors in the dimer binding energy before applying the CP correction.

The scatter plots in Figs.~\ref{fig:Ed_scatter} and \ref{fig:Ed2_scatter} show these results in more detail, by comparing either a NAO or a lower-cutoff PW basis with the converged PW basis, and plotting the energy difference between each pair of snapshots; both densities are included in the plot. Low-accuracy PW calculations give a large scatter of results, while all the NAO bases considered give a tight clustering around the diagonal, as do the higher PW cutoffs. Indeed, the NAO bases show no spurious outliers at all, and, equally importantly, no obvious systematic trend away from the diagonal that might cause a significant difference in the region of configuration space explored during a MD run.

\begin{figure}
\begin{center}
{\small{\begin{picture}(5040.00,3528.00)%
    \gdef\gplbacktext{}%
    \gdef\gplfronttext{}%
    \gplgaddtomacro\gplbacktext{%
      \csname LTb\endcsname%
      \put(1210,704){\makebox(0,0)[r]{\strut{} 0.001}}%
      \put(1210,1344){\makebox(0,0)[r]{\strut{} 0.01}}%
      \put(1210,1984){\makebox(0,0)[r]{\strut{} 0.1}}%
      \put(1210,2623){\makebox(0,0)[r]{\strut{} 1}}%
      \put(1210,3263){\makebox(0,0)[r]{\strut{} 10}}%
      \put(1342,484){\makebox(0,0){\strut{} 0}}%
      \put(2002,484){\makebox(0,0){\strut{} 400}}%
      \put(2662,484){\makebox(0,0){\strut{} 800}}%
      \put(3323,484){\makebox(0,0){\strut{} 1200}}%
      \put(3983,484){\makebox(0,0){\strut{} 1600}}%
      \put(4643,484){\makebox(0,0){\strut{} 2000}}%
      \put(176,1983){\rotatebox{-270}{\makebox(0,0){\strut{}RMS error (eV/\AA)}}}%
      \put(2992,154){\makebox(0,0){\strut{}PW cutoff (eV)}}%
      \put(2085,2205){\makebox(0,0)[r]{\strut{}$\mathrm{d}\zeta+\mathrm{p}$}}%
      \put(2085,1840){\makebox(0,0)[r]{\strut{}$\mathrm{t}\zeta+\mathrm{p}$}}%
      \put(2085,1481){\makebox(0,0)[r]{\strut{}$\mathrm{q}\zeta+\mathrm{dp}$}}%
    }%
    \gplgaddtomacro\gplfronttext{%
      \csname LTb\endcsname%
      \put(3656,3090){\makebox(0,0)[r]{\strut{}1.20 $\mathrm{g}/\mathrm{cm}^3$}}%
      \csname LTb\endcsname%
      \put(3656,2870){\makebox(0,0)[r]{\strut{}1.00 $\mathrm{g}/\mathrm{cm}^3$}}%
    }%
    \gplbacktext
    \put(0,0){\includegraphics{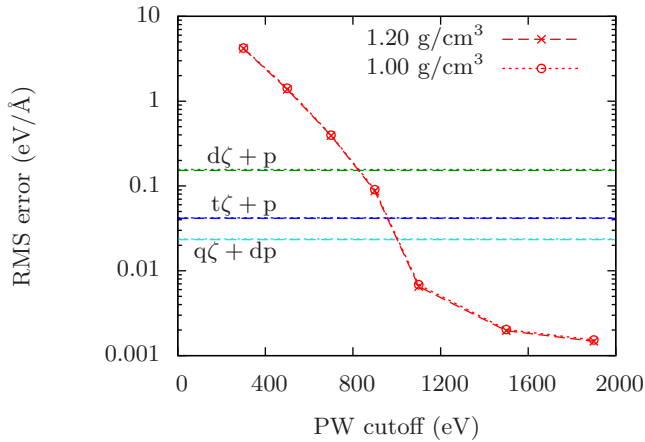}}%
    \gplfronttext
  \end{picture}}}
\caption{RMS error in the magnitude of the forces on all ions for 100 snapshots at each density. The points indicate PW calculations at different cutoffs, and the dashed horizontal lines show the value for three NAO basis sets.}
\label{fig:F_conv}
\end{center}
\end{figure}

Fig.~\ref{fig:F_conv} shows the RMS error in ionic forces. The results are very similar to those already discussed for energy differences, with the three new NAO bases giving errors equivalent to PW cutoffs of 830~eV, 960~eV, and 1000~eV, respectively. Differences between O and H ions are negligible (see Table~\ref{table:water_summary}). The table also reports results for the RMS angle between the forces obtained with the NAO basis and the converged PW one; these range from $\sim$14$^{\circ}$ for the \DZP\ basis to $\sim$1$^{\circ}$ for the \QZDP\ one. It is noteworthy that the \DZPmv\ basis performs better than the newly parametrized version in terms of force magnitudes and angles; this explains the relatively accurate relaxed geometries obtained for ice (see Table~\ref{table:ice_summary}).

\begin{figure}
\begin{center}
{\small{\begin{picture}(5040.00,3528.00)%
    \gdef\gplbacktext{}%
    \gdef\gplfronttext{}%
    \gplgaddtomacro\gplbacktext{%
      \csname LTb\endcsname%
      \put(1078,704){\makebox(0,0)[r]{\strut{}-1600}}%
      \put(1078,1216){\makebox(0,0)[r]{\strut{}-1200}}%
      \put(1078,1728){\makebox(0,0)[r]{\strut{}-800}}%
      \put(1078,2239){\makebox(0,0)[r]{\strut{}-400}}%
      \put(1078,2751){\makebox(0,0)[r]{\strut{} 0}}%
      \put(1078,3263){\makebox(0,0)[r]{\strut{} 400}}%
      \put(1210,484){\makebox(0,0){\strut{} 0}}%
      \put(1897,484){\makebox(0,0){\strut{} 400}}%
      \put(2583,484){\makebox(0,0){\strut{} 800}}%
      \put(3270,484){\makebox(0,0){\strut{} 1200}}%
      \put(3956,484){\makebox(0,0){\strut{} 1600}}%
      \put(4643,484){\makebox(0,0){\strut{} 2000}}%
      \csname LTb\endcsname%
      \put(176,1983){\rotatebox{-270}{\makebox(0,0){\strut{}Average pressure shift (kbar)}}}%
      \put(2926,154){\makebox(0,0){\strut{}PW cutoff (eV)}}%
    }%
    \gplgaddtomacro\gplfronttext{%
      \csname LTb\endcsname%
      \put(3656,1537){\makebox(0,0)[r]{\strut{}1.20 $\mathrm{g}/\mathrm{cm}^3$}}%
      \csname LTb\endcsname%
      \put(3656,1317){\makebox(0,0)[r]{\strut{}1.00 $\mathrm{g}/\mathrm{cm}^3$}}%
      \csname LTb\endcsname%
      \put(3656,1097){\makebox(0,0)[r]{\strut{}1.20 $\mathrm{g}/\mathrm{cm}^3$ (MV)}}%
      \csname LTb\endcsname%
      \put(3656,877){\makebox(0,0)[r]{\strut{}1.00 $\mathrm{g}/\mathrm{cm}^3$ (MV)}}%
    }%
    \gplbacktext
    \put(0,0){\includegraphics{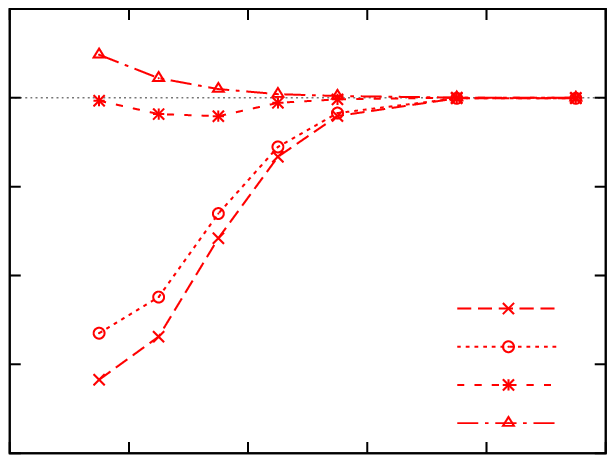}}%
    \gplfronttext
  \end{picture}}{\begin{picture}(2520.00,3528.00)%
    \gdef\gplbacktext{}%
    \gdef\gplfronttext{}%
    \gplgaddtomacro\gplbacktext{%
      \csname LTb\endcsname%
      \put(594,704){\makebox(0,0)[r]{\strut{}-10}}%
      \put(594,1344){\makebox(0,0)[r]{\strut{}-5}}%
      \put(594,1984){\makebox(0,0)[r]{\strut{} 0}}%
      \put(594,2623){\makebox(0,0)[r]{\strut{} 5}}%
      \put(594,3263){\makebox(0,0)[r]{\strut{} 10}}%
      \put(726,484){\makebox(0,0){\strut{} 1000}}%
      \put(1425,484){\makebox(0,0){\strut{} 1500}}%
      \put(2123,484){\makebox(0,0){\strut{} 2000}}%
      \csname LTb\endcsname%
      \put(176,1983){\rotatebox{-270}{\makebox(0,0){\strut{}}}}%
      \put(1424,154){\makebox(0,0){\strut{}}}%
      \put(1983,2621){\makebox(0,0)[r]{\strut{}$\mathrm{d}\zeta+\mathrm{p}$}}%
      \put(1983,2307){\makebox(0,0)[r]{\strut{}$\mathrm{t}\zeta+\mathrm{p}$}}%
      \put(1983,1808){\makebox(0,0)[r]{\strut{}$\mathrm{q}\zeta+\mathrm{dp}$}}%
    }%
    \gplgaddtomacro\gplfronttext{%
    }%
    \gplbacktext
    \put(0,0){\includegraphics{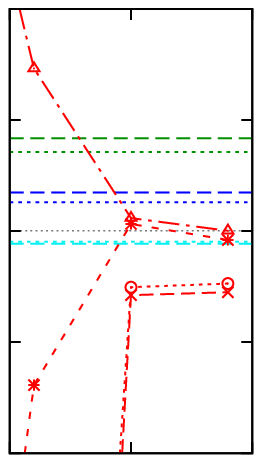}}%
    \gplfronttext
  \end{picture}}}
\caption{The average shift in pressure for 100 snapshots at each density. The points indicate PW calculations at different cutoffs (both with and without the MV correction for finite PW cutoff), and the dashed horizontal lines show the value for three NAO basis sets with no correction applied.}
\label{fig:P_conv}
\end{center}
\end{figure}

\begin{figure}
\begin{center}
{\small{\begin{picture}(5040.00,3528.00)%
    \gdef\gplbacktext{}%
    \gdef\gplfronttext{}%
    \gplgaddtomacro\gplbacktext{%
      \csname LTb\endcsname%
      \put(682,704){\makebox(0,0)[r]{\strut{}-2}}%
      \put(682,1216){\makebox(0,0)[r]{\strut{} 0}}%
      \put(682,1728){\makebox(0,0)[r]{\strut{} 2}}%
      \put(682,2239){\makebox(0,0)[r]{\strut{} 4}}%
      \put(682,2751){\makebox(0,0)[r]{\strut{} 6}}%
      \put(682,3263){\makebox(0,0)[r]{\strut{} 8}}%
      \put(814,484){\makebox(0,0){\strut{}-20}}%
      \put(1293,484){\makebox(0,0){\strut{}-15}}%
      \put(1771,484){\makebox(0,0){\strut{}-10}}%
      \put(2250,484){\makebox(0,0){\strut{}-5}}%
      \put(2729,484){\makebox(0,0){\strut{} 0}}%
      \put(3207,484){\makebox(0,0){\strut{} 5}}%
      \put(3686,484){\makebox(0,0){\strut{} 10}}%
      \put(4164,484){\makebox(0,0){\strut{} 15}}%
      \put(4643,484){\makebox(0,0){\strut{} 20}}%
      \csname LTb\endcsname%
      \put(176,1983){\rotatebox{-270}{\makebox(0,0){\strut{}Basis set error (kbar)}}}%
      \put(2728,154){\makebox(0,0){\strut{}P (kbar)}}%
    }%
    \gplgaddtomacro\gplfronttext{%
      \csname LTb\endcsname%
      \put(1870,3090){\makebox(0,0)[r]{\strut{}$\mathrm{d}\zeta+\mathrm{p}$}}%
      \csname LTb\endcsname%
      \put(1870,2870){\makebox(0,0)[r]{\strut{}$\mathrm{t}\zeta+\mathrm{p}$}}%
      \csname LTb\endcsname%
      \put(1870,2650){\makebox(0,0)[r]{\strut{}$\mathrm{q}\zeta+\mathrm{dp}$}}%
    }%
    \gplbacktext
    \put(0,0){\includegraphics{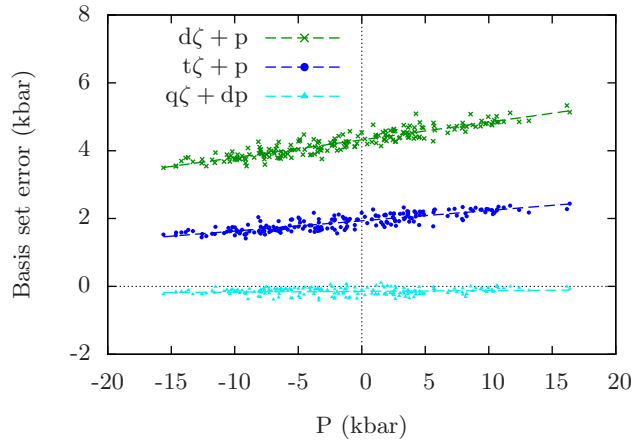}}%
    \gplfronttext
  \end{picture}}}
\caption{Error in NAO basis set pressures with respect to converged PW pressures for all 200 snapshots. The dashed lines show the linear fits used for the basis set correction in Fig.~\ref{fig:Pd_conv} and Table~\ref{table:water_summary}.}
\label{fig:P_fit}
\end{center}
\end{figure}

Finally, we examine the accuracy of our bases for calculating cell pressures. It is well-known that PW calculations suffer in this respect from the fact that the kinetic energy cutoff effectively changes with an infinitesimal change in volume (since the calculation of stresses is implicitly performed at constant PW number), leading to a spurious tensile stress. In order to elimate this, the correction by Meade and Vanderbilt~\cite{Meade1989} (here referred to as MV) is generally applied. The MV correction is given by
\begin{equation}
\frac{2}{3V} \frac{\partial E (E^\mathrm{cut},V)}{\partial \ln E^\mathrm{cut}},
\end{equation}
where $E$ is the total energy, $V$ the cell volume, and $E^\mathrm{cut}$ the kinetic energy cutoff. In contrast, the only systematic error introduced by NAO basis sets in the calculation of stresses is that from BSSE.

Fig.~\ref{fig:P_conv} shows the average shift in pressure with respect to the converged calculations. Even after applying the MV correction term, the PW calculations give very large shifts, on the order of 10--100~kbar. The NAO shifts, instead, are of no more than 4~kbar (for the \DZP\ basis). The PW convergence also varies significantly with density, which is not the case for the NAOs.

We can use our results to define a simple and effective correction for cell pressures calculated with NAOs. This is illustrated in Fig.~\ref{fig:P_fit}, showing a scatter plot of NAO pressures with respect to the converged PW result (note that we are plotting the error in the NAO value against the absolute value from PWs). Two trends can be clearly observed: a constant shift to higher pressures (i.e., the same result as shown in Fig.~\ref{fig:P_conv}), and an increase in the stiffness of the system; both of these reduce with basis size, and are eliminated almost completely for the \QZDP\ basis. Therefore, by performing a linear fit to the data points for a given basis we obtain an expression for correcting pressure values in a large range. It is important to note that the figure includes data points at both densities, which show the same trend; the correction can therefore also be used independently of density, at least within the range considered.

\begin{figure}
\begin{center}
{\small{\begin{picture}(5040.00,3528.00)%
    \gdef\gplbacktext{}%
    \gdef\gplfronttext{}%
    \gplgaddtomacro\gplbacktext{%
      \csname LTb\endcsname%
      \put(1210,704){\makebox(0,0)[r]{\strut{} 0.001}}%
      \put(1210,1344){\makebox(0,0)[r]{\strut{} 0.01}}%
      \put(1210,1984){\makebox(0,0)[r]{\strut{} 0.1}}%
      \put(1210,2623){\makebox(0,0)[r]{\strut{} 1}}%
      \put(1210,3263){\makebox(0,0)[r]{\strut{} 10}}%
      \put(1342,484){\makebox(0,0){\strut{} 0}}%
      \put(2002,484){\makebox(0,0){\strut{} 400}}%
      \put(2662,484){\makebox(0,0){\strut{} 800}}%
      \put(3323,484){\makebox(0,0){\strut{} 1200}}%
      \put(3983,484){\makebox(0,0){\strut{} 1600}}%
      \put(4643,484){\makebox(0,0){\strut{} 2000}}%
      \put(176,1983){\rotatebox{-270}{\makebox(0,0){\strut{}RMS error (kbar)}}}%
      \put(2992,154){\makebox(0,0){\strut{}PW cutoff (eV)}}%
      \put(2085,2453){\makebox(0,0)[r]{\strut{}$\mathrm{d}\zeta+\mathrm{p}$}}%
      \put(2085,2270){\makebox(0,0)[r]{\strut{}$\mathrm{t}\zeta+\mathrm{p}$}}%
      \put(2085,1894){\makebox(0,0)[r]{\strut{}$\mathrm{q}\zeta+\mathrm{dp}$}}%
      \put(3653,2339){\makebox(0,0)[l]{\strut{}(linear fit)}}%
    }%
    \gplgaddtomacro\gplfronttext{%
      \csname LTb\endcsname%
      \put(3656,3090){\makebox(0,0)[r]{\strut{}1.20 $\mathrm{g}/\mathrm{cm}^3$}}%
      \csname LTb\endcsname%
      \put(3656,2870){\makebox(0,0)[r]{\strut{}1.00 $\mathrm{g}/\mathrm{cm}^3$}}%
    }%
    \gplbacktext
    \put(0,0){\includegraphics{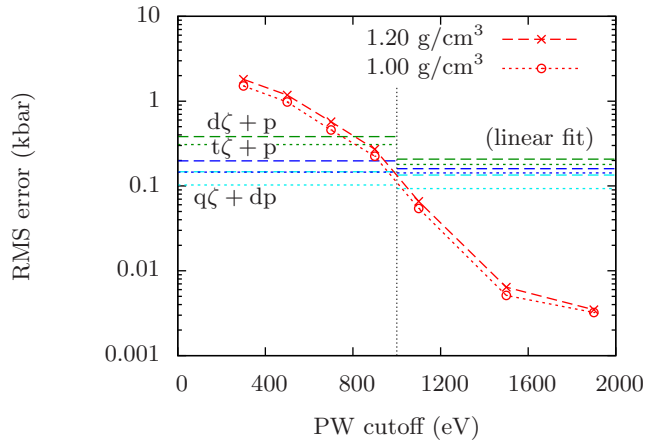}}%
    \gplfronttext
  \end{picture}}}
\caption{RMS error in differences in pressures between 100 snapshots at each density. The points indicate PW calculations at different cutoffs, and the dashed horizontal lines show the value for three NAO basis sets (uncorrected, on the left side of the plot, and with the correction explained in the text on the right side).}
\label{fig:Pd_conv}
\end{center}
\end{figure}

The effect of our correction for NAO calculations is given in Table~\ref{table:water_summary} for average pressure shifts and RMS errors in pressure differences. The correction reduces the average shift to very small values ($\sim$0.5~kbar) for all bases, the remaining error being due to the small discrepancy between the uncorrected shifts at the two different densities shown in Fig.~\ref{fig:P_conv}. Pressure differences between snapshots are generally of less interest for MD simulations than absolute values; nevertheless, they can be used as an additional measure of the quality of the basis. The comparison with PWs is shown in Fig.~\ref{fig:Pd_conv}, both for the uncorrected and corrected NAO results (the MV correction for PWs has no effect on pressure differences). The three uncorrected NAO bases give errors equivalent to PW cutoffs of approximately 810~eV, 950~eV, and 1000~eV, respectively, similarly to the values reported previously for other quantities of interest. After the correction is applied, the error for all bases is reduced to $\sim$0.2~kbar. Once again, we note the peculiarity of the \DZPmv\ basis (used in Ref.~\cite{water_arxiv}), which features a very small average pressure shift without the need for a correction, while the RMS error in pressure differences follows the trend of the other bases.

\begin{table}
\caption{Summary of results for liquid water. Listed are the average shift/RMS error for all snapshots at each density for various quantites: total energies ($E$), total pressures ($P$), energy differences ($\Delta E$), pressure differences ($\Delta$), magnitudes of ionic forces ($|\mathbf{F}|$), and angles between the SIESTA NAO force ($\mathbf{F}$) and the converged PW force ($\mathbf{F}_0$). In all cases the reference values are taken from ABINIT PW calculations at 2700~eV cutoff.}
\begin{center}
{\footnotesize
\lineup
\begin{tabular}{@{}lcccccccc}
\br
                                                                                                                                                                                               &&&                   & \multicolumn{5}{c}{Basis set} \\ \noalign{\smallskip}\cline{5-9}\noalign{\smallskip}
Average shift                                                                                                                                                                                  &&& $\rho$ (g/cm$^3$) & \DZPmv  & \DZP      & \TZPmv  & \TZP    & \QZDP   \\
\mr
\multicolumn{3}{l}{\multirow{2}{*}{$E$/molecule (eV)}}                                                                                                                                           & 1.20              &  \m0.28 &  \m\00.19 &  \m0.15 &  \m0.12 &  \m0.08 \\
                                                                                                                                                 &&                                              & 1.00              &  \m0.29 &  \m\00.21 &  \m0.16 &  \m0.12 &  \m0.08 \\
\mr
\multirow{4}{*}{$P$ (kbar)}                                                                                                                       & \multicolumn{2}{c}{\multirow{2}{*}{uncorr.}} & 1.20              &  \m0.06 &  \m\04.18 &  \m2.48 &  \m1.73 & $-$0.57 \\
                                                                                                                                                 &&                                              & 1.00              & $-$0.89 &  \m\03.56 &  \m1.99 &  \m1.30 & $-$0.48 \\
                                                                                                                                                  & \multicolumn{2}{c}{\multirow{2}{*}{corr.}}   & 1.20              & $-$0.27 & \0$-$0.55 & $-$0.43 & $-$0.37 & $-$0.43 \\
                                                                                                                                                 &&                                              & 1.00              & $-$0.48 & \0$-$0.65 & $-$0.53 & $-$0.50 & $-$0.32 \\
\mr
RMS error \\
\mr
\multicolumn{3}{l}{\multirow{2}{*}{$\Delta E$/molecule (meV)}}                                                                                                                                   & 1.20              &  \m1.86 &  \m\01.90 &  \m1.32 &  \m0.93 &  \m0.47 \\
                                                                                                                                                 &&                                              & 1.00              &  \m1.54 &  \m\01.85 &  \m1.44 &  \m0.94 &  \m0.43 \\
\mr
\multirow{4}{*}{$\Delta P$ (kbar)}                                                                                                                & \multicolumn{2}{c}{\multirow{2}{*}{uncorr.}} & 1.20              &  \m0.41 &  \m\00.38 &  \m0.27 &  \m0.20 &  \m0.15 \\
                                                                                                                                                 &&                                              & 1.00              &  \m0.39 &  \m\00.31 &  \m0.22 &  \m0.15 &  \m0.10 \\
                                                                                                                                                  & \multicolumn{2}{c}{\multirow{2}{*}{corr.}}   & 1.20              &  \m0.33 &  \m\00.21 &  \m0.14 &  \m0.16 &  \m0.13 \\
                                                                                                                                                 &&                                              & 1.00              &  \m0.24 &  \m\00.18 &  \m0.14 &  \m0.14 &  \m0.09 \\
\mr
\multicolumn{2}{l}{\multirow{4}{*}{$|\mathbf{F}|$ (eV/\AA)}}                                                                                      & \multirow{2}{*}{O}                           & 1.20              &  \m0.11 &  \m\00.17 &  \m0.07 &  \m0.05 &  \m0.03 \\
                                                                                                                                                 &&                                              & 1.00              &  \m0.11 &  \m\00.17 &  \m0.07 &  \m0.05 &  \m0.03 \\
                                                                                                                                                 && \multirow{2}{*}{H}                           & 1.20              &  \m0.07 &  \m\00.15 &  \m0.09 &  \m0.04 &  \m0.02 \\
                                                                                                                                                 &&                                              & 1.00              &  \m0.07 &  \m\00.15 &  \m0.09 &  \m0.04 &  \m0.02 \\
\mr
\multicolumn{2}{c}{\multirow{4}{*}{$\cos^{-1} \left \{ \frac{\mathbf{F} \cdot \mathbf{F_0}}{|\mathbf{F}||\mathbf{F_0}|} \right \}$ ($^{\circ}$)}} & \multirow{2}{*}{O}                           & 1.20              &  \m8.27 &   \m13.83 &  \m5.41 &  \m4.33 &  \m0.98 \\
                                                                                                                                                 &&                                              & 1.00              &  \m7.56 &   \m13.40 &  \m5.00 &  \m4.33 &  \m1.04 \\
                                                                                                                                                 && \multirow{2}{*}{H}                           & 1.20              &  \m7.57 &   \m15.11 &  \m8.91 &  \m3.88 &  \m1.77 \\
                                                                                                                                                 &&                                              & 1.00              &  \m7.34 &   \m15.01 &  \m8.77 &  \m4.07 &  \m1.67 \\
\br
\end{tabular}
\label{table:water_summary}
}
\end{center}
\end{table}

\section{Conclusions}

In this paper, we have described the development and testing of finite-range NAO basis sets for water-based systems. We have discussed the general strategy employed for creating basis sets of increasing size and accuracy, and have proposed the use of a screened Coulomb confinement potential for shaping the polarization orbitals, in order to achieve a good agreement with the higher angular momentum shells of the free atom without needing to extend the confinement radius beyond what is physically useful for condensed matter systems.

We have presented 13 new bases for the water molecule, ranging from \DZP\ to \QZTPDP. The full list of parameters needed to recreate and use them are given in the Supplementary Material (we provide these instead of the numerical radial functions themselves so as to minimize the dependence on specific pseudopotentials).

In order to perform rigorous tests of the accuracy of our bases, we use auxiliary PW calculations at different kinetic energy cutoffs. We can therefore compare the results obtained with NAOs with the PW ones at a very large (essentially fully converged) cutoff, and also equate the accuracy of the various NAO bases with PW bases at specific cutoffs. This is done using two different DFT codes (SIESTA for NAOs, ABINIT for PWs) with the same pseudopotentials and Kleinman-Bylander factorization; even so, small algorithmic differences between the codes will affect the comparison with the NAO bases of highest accuracy, slightly underestimating their performance with respect to PWs for quantities such as energy differences.

The results for our tests on a variety of molecular and condensed systems show the transferability and accuracy of the bases. In particular, there is a good level of consistency both between different systems and different properties of the same system when comparing to the performance of PW calculations at finite cutoff: errors in total energies for the \DZP, \TZP, and \QZDP\ bases are on the order of those for cutoffs of 1100~eV, 1200~eV, and 1300~eV, respectively, while errors in energy differences, ionic forces, and pressure differences are on the order of those for cutoffs of 800~eV, 900~eV, and 1000~eV. However, it is important to remember that the two types of bases are not at all equivalent: this is clearly demonstrated by calculations of absolute pressure, which show NAOs to naturally give much smaller errors for this quantity than PWs. We have also proposed a simple correction for further reducing errors in absolute pressures and pressure differences for NAO calculations, based on a linear fit to the data obtained from 200 liquid water snapshots at two densities; this can be used, e.g., to correct average pressures obtained from NVE AIMD simulations.

\ack

This work was partly funded by grants FIS2009-12721 and FIS2012-37549 from the Spanish Ministry of Science. MVFS acknowledges a DOE Early Career Award No. DE-SC0003871. We thank Javier Junquera for support with the translation of pseudopotentials between SIESTA and ABINIT. The calculations were performed on the following HPC clusters: kroketa (CIC nanoGUNE, Spain), arina (Universidad del Pa\'{i}s Vasco/Euskal Herriko Unibertsitatea, Spain), magerit (CeSViMa, Universidad Polit\'{e}cnica de Madrid, Spain). We thank the RES--Red Espa\~nola de Supercomputaci\'{o}n for access to magerit. SGIker (UPV/EHU, MICINN, GV/EJ, ERDF and ESF) support is gratefully acknowledged.

\end{document}